\documentclass{aa}  

\usepackage{graphicx}
\usepackage{subfiles}
\usepackage{txfonts}
\usepackage{todonotes}

\usepackage{xspace}
\newcommand{\vac}{\texttt{MPI-AMRVAC}\xspace}

\usepackage{hyperref}

\begin{document}

   \title{Radiation-Hydrodynamics with \vac}

   \subtitle{Flux-Limited Diffusion}

   \author{N. Moens\inst{1}
          \and
          J.O. Sundqvist\inst{1}
          \and
          I. El Mellah\inst{1,2,4}
          \and 
          L. Poniatowski\inst{1}
          \and
          J. Teunissen\inst{2,3}
          \and
          R. Keppens\inst{2}
          }

   \institute{Instituut voor Sterrenkunde, KU Leuven,
              Celestijnenlaan 200D, 3001 Leuven, Belgium,\\
              \email{nicolas.moens@kuleuven.be}
         \and
             Centre for mathematical Plasma Astrophysics, Department of Mathematics, KU Leuven,
             Celestijnenlaan 200B, 3001 Leuven, Belgium
        \and 
             Centrum Wiskunde \& Informatica, 
             PO Box 94079, 1090 GB Amsterdam, The Netherlands
        \and 
            Institut de Planétologie et d’Astrophysique de Grenoble (IPAG), Université Grenoble Alpes, 38058 Grenoble Cedex 9, France
            }

   \date{Received ...; accepted ...}

 
  \abstract
   {Radiation controls the dynamics and energetics of many astrophysical environments. To capture the coupling between the radiation and matter, however, is often a physically complex and computationally expensive endeavour.}
   {We develop a numerical tool to perform radiation-hydrodynamics simulations in various configurations at an affordable cost.}
   {We build upon the finite volume code \texttt{MPI-AMRVAC} to solve the equations of hydrodynamics on multi-dimensional adaptive meshes and introduce a new module to handle the coupling with radiation. A non-equilibrium, flux-limiting diffusion approximation is used to close the radiation momentum and energy equations. The time-dependent radiation energy equation is then solved within a flexible framework, accounting fully for radiation forces and work terms and further allowing the user to adopt a variety of descriptions for the radiation-matter interaction terms (the 'opacities').} 
   {We validate the radiation module on a set of standard testcases for which different terms of the radiative energy equation predominate. As a preliminary application to a scientific case, we calculate spherically symmetric models of the radiation-driven and optically thick supersonic outflows from massive Wolf-Rayet stars. This also demonstrates our code's flexibility, as the illustrated simulation combines opacities typically used in static stellar structure models with a parametrised form for the enhanced line-opacity expected in supersonic flows.}
   {This new module provides a convenient and versatile tool to perform multi-dimensional and high resolution radiative-hydrodynamics simulations in optically thick environments with the \texttt{MPI-AMRVAC} code. The code is ready to be used for a variety of astrophysical applications, where a first target for us will be multi-dimensional simulations of stellar outflows from Wolf-Rayet stars.}


   \maketitle
%


\section{Introduction} \label{sec:introduction}


 








 

In many astrophysical environments, radiation plays an important role in the system's total energy and/or momentum budget. Some selected examples involve: solar convection simulations where radiation controls the heating and cooling of the photosphere \citep{Stein1998}, wind outflows from massive stars where a strong radiation force lifts material off the stellar surface \citep{CAK1975}, evacuation of massive star-forming discs where radiative ablation may control the stellar upper mass limit \citep{Kee2019}, accretion flows around \citep{Jiang2019}, and disc winds from \citep{Proga2004} supermassive black holes, where radiation often is the dominant player controlling the 
energetics and dynamics. 

Implementation of suitable radiation modules for dynamical simulations, however, is a very difficult task in general \citep{Mihalas1984a, Castor2004}. Because of this, a rather wide range of different approaches have been taken depending on the specific application. For example, in the aforementioned convection simulations of solar-type stars, the heating and cooling terms are computed directly from solutions to the time-independent radiative transfer equation (RTE) (using a sort of "frozen-in" approach, e.g. \citealt{Stein1998}), and the effect on the momentum balance is neglected. Similar considerations apply for solar coronal and wind simulations, however here the radiative cooling term is often approximated by assuming an optically thin approximation \citep[e.g.,][]{Schure2009}. On the other hand, the winds of hot, massive stars are driven by momentum transfer from the radiation field to the gas. For such simulations, the force due to spectral lines is critical and very elaborate radiation transport methods have meanwhile been developed for steady outflows \citep{Sander2017, Sundqvist2019}, whereas time-dependent simulations typically rely on various distribution-function approaches \citep{Owocki1988, Sundqvist2018}. 

For general attempts to solve the full time-dependent radiation momentum and energy equations, two important complicating factors are {\it i)} how to obtain a suitable closure relation for the radiation equations and {\it ii)} how to compute the radiation-matter interaction terms (the opacities) in a supersonic flow. Concerning closure relation, one may attempt to obtain closure by a variable Eddington tensor (VET) computed directly from the radiative transfer equation, using for example a short characteristics scheme \citep{Jiang2012}. Alternatively, various analytic closure relations are often applied, such as flux limited diffusion (FLD) \citep{Turner12001, Krumholz2006}, first moment (M1) \citep{Skinner2013, Bloch21}, or a combination of both \citep{Mignon2020}. Concerning the radiation-matter interaction terms, the central issue is that it is only in the frame co-moving with the fluid (the CMF) that opacities normally are isotropic. As such, only in this frame suitable mean opacities can be defined without performing complicated double integrals of the specific intensity and opacity over frequency and angle. On the other hand, the radiative transfer equation itself is significantly more complicated in the CMF than in the laboratory frame (see discussions in \citealt{Mihalas1984a, Castor2004}). As such, some trade-offs must typically be done, where one popular method is to compute the radiation variables in the laboratory frame and then via a first order expansion still treat the opacities in the CMF \citep{Mihalas1982, Lowrie1999, Jiang2012}. A main drawback with this 
"mixed-frame" formulation, however, is that it is not suitable for computing line opacities in a supersonic flow, since the first order expansion upon which it is based does not apply for the rapidly changing opacities of spectral lines. 

\vac\footnote{\href{http://amrvac.org/}{http://amrvac.org/}} is a modern, MPI-parallelized computer code, aimed at solving partial differential equations (PDEs) on an adaptive, block structured quadtree/octree mesh \citep{Xia2017, Keppens2021}. Originally, the code focused on the equations of (magneto-) hydrodynamics, but recently the code has been expanded to solve general hyperbolic, parabolic and elliptic PDEs. The code uses a finite volume method to solve hyperbolic advection equations in 1-3 dimensions and includes a large number of schemes and limiters for flux reconstruction, time discretisation methods and refinement strategies. In recent years, multiple new physics modules have been added to \vac, such as descriptions of viscosity, conduction, and dust dynamics \citep{Porth14, Xia2017}. One aspect that has been missing, however, is a general description of the dynamical effects of a radiation field. Thus far, only effects of optically thin radiative cooling have been implemented \citep{Schure2009, vanMarle11}. In this paper, we take a first step toward a more general description, implementing a non-equilibrium FLD method into \vac. For many of our targeted applications line-opacity is crucial; as such we formulate the FLD equations in the CMF (e.g., Turner \& Stone 2001) and also present a simplified way (based on Poniatowski et al. 2021) of accounting for supersonic line-opacities within the formalism. Recently, \vac has been expanded to include the possibility to solve elliptic PDEs \citep{Keppens2021} using a geometric multigrid library \citep{Teunissen2019}. This has been crucial for a proper treatment of the diffusive term in the FLD equations, ensuring a stable solution by applying an implicit method.  

The paper is structured as follows: In section \ref{sec:EquationsRHD} we introduce the radiative-hydrodynamics (RHD) equations and we describe the non-equilibrium FLD method for obtaining closure. The different aspects of the implementation in the \vac code are explained in section \ref{sec:NumImpl}. Section \ref{sec:Testcases} is devoted to benchmark tests for the newly developed code,  
and in section \ref{sec:StellarAppl} we present a first research application in the form of an optically thick radiation-driven wind outflow from a classical Wolf-Rayet star. Section \ref{sec:Conclusions} summarizes and discusses the paper and provides an outlook. 

\section{Equations of RHD} \label{sec:EquationsRHD}
Including source terms due to radiation, the hydrodynamical conservation equations of mass, momentum, and energy are: 
\begin{align}
\partial_t \rho + \nabla \cdot (\rho \vec{v}) &= 0, \label{eq:hd_rho}\\
\partial_t (\rho \vec{v}) + \nabla \cdot (\vec{v} \rho \vec{v} + p\vec{I}) &= \vec{f}_r, \label{eq:hd_mom} \\
 \partial_t e +\nabla \cdot (e \vec{v} + p \vec{v}) &=  \vec{v} \cdot \vec{f}_r + \dot{q}, \label{eq:hd_e}
\end{align}
where $\rho$ is the gas density, $\vec{v}$ is the gas velocity, $e$ the total gas energy density (internal plus kinetic), $p$ is the gas pressure and $\vec{I}$ is the identity matrix. On the right hand side, ${\vec f}_r$ is the radiation force density exerted on the matter
and $\dot{q}$ the heating/cooling term due to the radiation. These equations are supplemented by the expression for gas energy $e$ and the ideal gas law:
\begin{align}
e &= \frac{p}{(\gamma - 1)} + \frac{\rho v^2}{2}, \label{eq:hd_gaslaw}\\
p &= \frac{k_b T_g}{m_p \mu} \rho. 
\end{align}
Here, $\gamma$ is the adiabatic index, $k_b$ is the Boltzmann constant, $T_g$ is the gas temperature, $m_p$ is the proton mass and $\mu$ is the mean molecular weight of the gas particles. As described below, the terms $\vec{f}_r$ and $\dot{q}$ can be computed by considering the energy and momentum equations for the radiation. We note further that the above equations neglect effects of gravity and magnetic fields. However, the modular structure of \vac enables to seamlessly integrate the new FLD module described in this paper in the main branch of the code in order to use it in physical situations accounting for more physics.



\subsection{Radiation energy and momentum equations}

Conservative equations can be derived for the radiation field by taking angular moments of the radiative transfer equation for specific intensity $I_\nu$. However, when describing the transport of radiation using such moment equations, particular attention has to be paid to the reference frame of the radiation quantities. In an outside observer's reference frame (equivalent to the Eulerian or laboratory frame of reference), 
%
%
computation of $\dot{q}$ and $\vec{f}_r$ requires radiation-material interaction terms, where these involve an important angle-dependence induced by the Doppler shift. For moving fluids, this makes it necessary to always carry out double integrals over frequency and angle in order to obtain the correct coupling terms. These issues are avoided by transforming to a frame co-moving with the local velocity of the fluid. In this co-moving frame (CMF), the frequency-integrated radiation energy and momentum equations are \citep{Mihalas1984a, Castor2004}:
\begin{equation} 
 \label{Eq:0th_cmf} 
	\partial_t E +  \nabla \cdot (E \vec{v}) + \nabla \cdot \vec{F} + \vec{P} : \nabla \vec{v} = -\dot{q}, 
\end{equation} %
\begin{equation} 
 \label{Eq:1st_cmf} 
	\frac{1}{c^2} \left( \partial_t \vec{F}  + \nabla \cdot (\vec{F} \vec{v}) \right) +  \nabla \cdot \vec{P} = -\vec{f}_r,
\end{equation} 
where $E$, $\vec{F}$, and $\vec{P}$ are the frequency-integrated radiation energy density, flux vector, and pressure tensor, respectively, evaluated in the CMF.
%
%
%
%
These radiation equations are now somewhat more complex than in the observer's frame, as they contain extra terms stemming from the transformation (see Chapter 6 in \citealt{Castor2004} for the full transformation properties between the observer and co-moving frames). For example, the fourth term on the left hand side of the energy equation \eqref{Eq:0th_cmf} describes the dyadic product between the radiation pressure tensor and the gradient of the gas velocity vector. This term can be interpreted physically as the energy that leaves the radiation system when the radiation field accelerates and provides work, i.e. it is the radiation work term. In optically thick supersonic media, this physical effect can become critical as it may diminish the radiation flux, and so also the radiation force, significantly. As such, it is sometimes called "photon tiring" \citep{Owocki1997}. 
On the other hand, all material interaction terms in the CMF are in most situations isotropic, which  simplifies tremendously the evaluation of the right-hand-sides in equations 
\eqref{Eq:0th_cmf}-\eqref{Eq:1st_cmf}. These coupling terms in equations \eqref{eq:hd_mom}-\eqref{eq:hd_e} and \eqref{Eq:0th_cmf}-\eqref{Eq:1st_cmf} can now be written as: 

%
%
\begin{equation} 
	\dot{q} = 
	 c \kappa_{E} \rho E - 4 \pi \kappa_{B} \rho B, \label{eq:rhd_heatcool}
\end{equation} 
\begin{equation} 
	\vec{f}_r= \frac{\rho \kappa_{F} \vec{F}}{c}, \label{eq:rhd_force}
\end{equation} 
where $B = \int_0^{\infty} B_\nu d\nu = \sigma T_g^4/\pi$ is the frequency-integrated Planck function, with $\sigma$ the Stefan-Boltzmann constant, and the opacities $\kappa_B$ and $\kappa_E$ are mass absorption 
coefficients measured in $cm^2 / g$. Specifically, equations 
\eqref{eq:rhd_heatcool}-\eqref{eq:rhd_force} involve the Planck, energy density, and flux mean opacities:  

\begin{equation} 
	\kappa_{\rm P} \equiv \frac{ \int B_\nu \kappa_\nu d \nu } {B}, 
\end{equation} 
\begin{equation} 
	\kappa_{\rm E} \equiv \frac{ \int E_\nu \kappa_\nu d \nu } { E}, 
\end{equation} 
\begin{equation} 
	\kappa_{\rm F, i} \equiv \frac{ \int F_{\nu,i} \kappa_\nu d \nu}{F_i},  
\end{equation} 
%
%
which follows directly from considering the frequency-dependent form of the coupling terms 
described by equations \eqref{eq:rhd_heatcool}-\eqref{eq:rhd_force}. 
Here, $F_i$ is the $i^{th}$ component of the flux vector. For the sake of simplicity, we assume an isotropic flux mean opacity throughout this paper, though the method can be expanded for anisotropic opacities as well.
The quantities with subscript $_\nu$ are the frequency dependent radiation quantities (e.g. $E_\nu$ is the radiation energy density at frequency $\nu$).
Note that the above formulation does not necessarily mean the source functions must be Planckian. Indeed, the same expressions are found also for a model where emission and extinction coefficients have both thermal absorption ('a') and coherent scattering ('s') contributions, i.e. when $\kappa_\nu = \kappa_\nu^a + \kappa_\nu^s$. In such a situation, however, it is critical to keep in mind that although the flux mean then involves the total opacity $\kappa_\nu$, the energy and Planck means should be evaluated using only the thermal absorption part $\kappa_\nu^a$ (see eqn. 77 in Mihalas \& Mihalas 1984 on their page 336, and also the corresponding discussion on their page 472).

\subsection{Non-equilibrium FLD closure relation}
\label{sec:ileyk_FLD} 
An additional relation between $\vec{P}$ and $E$ is needed to close the radiation moment equations \eqref{Eq:0th_cmf} and \eqref{Eq:1st_cmf}. In general, this relation must be obtained from full solutions of the frequency and angle dependent transfer equation in different directions. However, realistic multi-frequency solutions to the radiative transfer equation in the CMF have thus far only been developed for 1D, steady-state media with a monotonic velocity field \citep[e.g.,][]{Hillier1998, Puls2020}. As such, analytic closure relations are often being used in practical radiation-hydrodynamics applications. Typically these analytic relations recover the correct equilibrium limit in the optically thick limit and then apply some appropriate "bridging law" for extension into the opposite optically thin streaming limit. 

In this paper we apply the so-called FLD approximation as our closure relation. Neglecting the first two terms in the radiation momentum equation, we have:
\begin{equation} 
    \nabla \cdot \vec{P} = \nabla \cdot \left(\vec{f} E\right) = - \frac{\kappa_F \rho}{c} \vec{F},  
\end{equation} 
for the Eddington tensor $\textit{\textbf{f} = \vec{P}/E}$. This invites us to write the radiation flux from Fick's diffusion law
(see also \citealt{Levermore1981}): 

\begin{equation}
        \vec{F} = -D \nabla E,
        \label{eq:ileyk_FLD}
\end{equation}
with $D$ a diffusion constant, which will depend on the local state of the gas and radiation field. Is is important to again note here that this diffusion approximation is only applicable for the CMF quantities of the radiation flux and energy density \citep{Mihalas1984a, Castor2004}.
%

In the limit of radiative diffusion, the energy density and pressure take their equilibrium values 
such that the scalar Eddington factor $f =1/3$ and the Eddington tensor becomes $\vec{f} = f\vec{I}$ with $\vec{I}$ the unit tensor. This is valid for very optically thick regions, where the photon mean-free path $\ell = 1/(\rho \kappa_F)$ is small compared to the typical length scales over which the state variables vary. Letting this length scale be the radiation energy density scale height $H_R = E/\nabla E$, we require $\ell \ll H_R$ in the radiative diffusion limit. This then results in a diffusion constant $D = c/(3 \kappa_F \rho)$. A basic issue with this diffusion approximation, however, is that the flux computed from it may exceed the physical limit $|\vec{F}| = c E$ in the opposite regime of freely streaming photons, where the diffusion constant will approach an arbitrarily large value as the local density and opacity approach zero. This suggests to introduce a bridging law that limits the flux in optically thin regions, while still recovering the optically thick limit. To this end, we apply the flux-limiter $\lambda$ suggested by \citet{Levermore1981} (see also \citealt{Turner12001}), writing:
\begin{equation} 
    f = \lambda + \lambda^2 R^2, \label{eq: fld_Edd_fac}
\end{equation} 
\begin{equation} 
    R = \frac{\ell}{H_R} = \frac{\nabla E}{\rho \kappa_F E},
\end{equation} 
and for the bridging law:
\begin{equation} 
    \lambda = \frac{2 +R}{6 + 3R + R^2} \label{eq:fld_lambda}. 
\end{equation} 
Clearly this relation recovers the diffusion limit since $\ell \ll H_R$ yields directly $\lambda \rightarrow 1/3$ and so $f \rightarrow 1/3$. Similarly in the opposite free streaming limit $\lambda \rightarrow 1/R$ such that $f \rightarrow 1$. We note that while we use this prescription throughout this paper, other variants have been suggested as well \citep[e.g.,][]{Minerbo1978}; in the 
\vac code, the user can readily switch between different flux limiters like the Levermore \citep{Levermore1981} or the Minerbo \citep{Minerbo1978} prescriptions. 

Using this flux-limiter within Fick's diffusion formulation, we can locally compute the co-moving radiation flux according to:
\begin{equation}
\vec{F} = \frac{-c\lambda}{\kappa_F \rho} \nabla E  \label{eq:fld_F}.    
\end{equation}
Here it can be seen that in the thin limit, when $\ell \gg H_R$, $\lambda \rightarrow 1/R$,  consequently $|\vec{F}| \rightarrow c E$ and causality is preserved.
Using then the corresponding Eddington tensor suggested by \citet{Turner12001}: 
\begin{equation}
        \vec{f} = \frac{1}{2}(1-f)\vec{I} + \frac{1}{2}(3f -1)\vec{\hat{n}}\vec{\hat{n}}, \label{eq:fld_Edd}
\end{equation}
we obtain also the radiation pressure tensor and so can omit equation \eqref{Eq:1st_cmf} entirely. Here, $\vec{\hat{n}}=\nabla E /|\nabla E|$ is the unit vector in the direction of the gradient of the radiation energy density (i.e. of the radiative flux). Thus, the only PDE that needs to be solved in order to advance the radiation subsystem is equation \eqref{Eq:0th_cmf}; this radiation energy equation describes the conservation of energy stored in the radiation field. 

In summary, the radiation energy density $E$ is integrated over time, the radiation flux $\vec{F}$ then calculated locally using the analytic FLD-approximation \eqref{eq:fld_F}, and the pressure tensor $\vec{P}$ obtained locally from $E$ using the analytic prescription for the Eddington tensor \eqref{eq:fld_Edd}.\\


For the radiative heating/cooling term $\dot{q}$ we further assume for the applications in this paper
that the Planck and energy density mean opacities are equal, $\kappa_P = \kappa_E = \kappa$, such that: 
\begin{equation} 
	\dot{q} = 
	 \rho \kappa  4 \sigma  \left(T_r^4 -  T_g^4 \right),  
\end{equation}     
where we have reformulated $E$ in terms of a radiation temperature $E \equiv a_r T_{r}^4$, where $a_r = 4 \sigma/c$ is the radiation constant. In the non-equilibrium FLD method applied here, $T_r$ can deviate from the gas temperature $T_g$, allowing for situations where the coupling between radiation and gas may be out of equilibrium.

\section{Numerical Implementation} \label{sec:NumImpl}
The RHD system of PDEs in the non-equilibrium FLD approximation consist of equations\,\eqref{eq:hd_rho}-\eqref{eq:hd_e} and equation\,\eqref{Eq:0th_cmf}. 
The conservative left hand side parts of the hydrodynamic equations \eqref{eq:hd_rho}-\eqref{eq:hd_e} are entirely hyperbolic and can be solved using the existing variety of shock capturing, high resolution finite volume solvers in \vac \citep{Xia2017}. However, when taking into account the coupling between the radiation field and its effects on the gas quantities, the system loses its purely hyperbolic property. This coupled system is now solved in an operator split manner, where different terms are added using different schemes, depending on the timescales of their effect. The advection term, $\nabla \cdot (E\vec{v})$ in equation\,\eqref{Eq:0th_cmf}, is handled using the same solvers already available in \vac for solving hyperbolic equations. Unless stated otherwise, in the test cases and applications presented in the following sections, we use the second-order accurate shock-capturing total variation diminishing Lax-Friedrichs (TVDLF) \citep{Toth1996} scheme and a Koren slope limiter \citep{Koren1993}. 
Since the effects of the radiative force $\vec{f}_{r}$ and its work $\vec{v} \cdot \vec{f}_{r}$ contribute on the same timescale as the advection with the gas velocity, namely the dynamical timescale, the corresponding source terms in the right hand side of the momentum and gas energy equations\,\eqref{eq:hd_mom} and \eqref{eq:hd_e} respectively are added explicitly, as is the photon tiring term $\vec{P}:\nabla \vec{v}$ in the radiation energy equation \eqref{Eq:0th_cmf}. 

On the other hand, the heating/cooling and radiation diffusion happen on timescales $\tau_{q}$ and $\tau_{\mathrm{diff}}$ which can be, depending on the regime, several orders of magnitude shorter than the dynamical timescale. Consequently, to ensure numerical stability, their corresponding terms, $\dot{q}$ and $\nabla \cdot \vec{F}$, in the gas and radiation energy equations (\eqref{eq:hd_e}, \eqref{Eq:0th_cmf}) are  computed following an implicit procedure. How these aforementioned source terms and methods are combined is explained in section \ref{subsec:OoO}. The rest of section \ref{sec:NumImpl} is then devoted to elaborating the different source terms separately. \\

The FLD module discussed below is implemented in Cartesian coordinates for 1D, 2D and 3D setups, thanks to the VAC-preprocessor. For illustrative purposes, we will only write out stencils corresponding to a 2D setup. Extension to 3D or reduction to a 1D setup is a matter of adding/subtracting an index in the equations below.

For now, in the application of the WR star in section \ref{sec:StellarAppl} a spherical correction is used to correct for the geometry. This is further described in Appendix A. Finally, the new module described in this paper can, in principle, also be directly applied to radiation-MHD calculations.

\subsection{Order of operations} \label{subsec:OoO}
In all generality, the PDEs of the FLD system, equations \eqref{eq:hd_rho}-\eqref{eq:hd_e} and \eqref{Eq:0th_cmf}, can be written as:

\begin{align}
    \partial_t \vec{u} &= \vec{A}_{\mathrm{adv}}(\vec{u}) + \vec{H}_{\mathrm{diff}}(\vec{u}) + \vec{S}_{\mathrm{ex}}(\vec{u}) + \vec{S}_{\mathrm{im}}(\vec{u}).
\end{align}
Here, $\vec{u}$ represents the vector of conservative variables $[\rho, \vec{v} \rho, e, E]$. $A_{adv}(\vec{u}) = -\nabla \cdot [\rho \vec{v}, \vec{v}\rho\vec{v} + p, e\vec{v} + p\vec{v},E\vec{v}]$ is the advection operator; $H_\mathrm{diff}(\vec{u})= [0,\vec{0},0,-\nabla \cdot \vec{F}]$ is the diffusion term; $\vec{S}_{\mathrm{ex}}(\vec{u}) = [0, \vec{f}_r, \vec{v}\cdot\vec{f}_r, -\vec{P}:\nabla \vec{v}]$ is a collection of source terms which will be handled explicitly: radiation force and work additions in momentum and total energy. Finally, $\vec{S}_{\mathrm{im}}(\vec{u}) = [0,\vec{0}, \dot{q}, -\dot{q}]$ are the entirely local source terms which are handled implicitly: cooling and heating. In our implementation in \vac, these four types of terms can be grouped in two operators: an Implicit operator and an Explicit operator:  \\

\begin{align}
    G_{ex}(\vec{u}) &= \vec{A}_{adv}(\vec{u}) + \vec{S}_{\mathrm{ex}}(\vec{u}) + \vec{S}_{\mathrm{im}}(\vec{u}), \\
    G_{im}(\vec{u}) &= \vec{H}_\mathrm{diff}(\vec{u}).
\end{align}
Note that although $G_{ex}$ is referred to as an explicit operator, it contains one local implicit step $\vec{S}_{\mathrm{im}}(\vec{u})$ which is further discussed in section\,\ref{subsec:HeatCool}. 
For the explicit operator $G_{ex}(\vec{u})$ and for illustrative purposes, we only show the simplest formulation. The scheme employed to advance the vector of conservative variables from $\vec{u}^{n}$ to $\vec{u}^{n+1}_{+}$ is: 

\begin{align}
    \vec{u}^{n+1}_*  &= \vec{u}^{n} + \Delta t \vec{S}_{\mathrm{ex}}(\vec{u}^n) \label{eq:G_expl_1},\\
    \vec{u}^{n+1}_{**} &= \vec{u}^{n+1}_* + \Delta t \vec{S}_{\mathrm{im}}(\vec{u}^{n+1}_{**}, \vec{u}^{n+1}_{*} ) \label{eq:G_expl_2},\\
    \vec{u}^{n+1}_{+}   &= \vec{u}^{n+1}_{**} + \Delta t  \vec{A}_{adv}(\vec{u}^{n+1}_{**}) \label{eq:G_expl_3},
\end{align}
%
%
where the states indexed with an asterisk or plus sign are intermediate states. As can be seen in equations \eqref{eq:G_expl_1}-\eqref{eq:G_expl_3}, the radiation force $\vec{f_r}$, its work $\vec{v}\cdot\vec{f_r}$ and the photon tiring term $\vec{P}:\nabla \vec{v}$ are added first. These are followed by the local implicit heating and cooling terms $\dot{q}$. Finally, the updated state is used in the advection of both the gas and radiation variables. This combination of terms described above can also be used when only advancing half a time step in for example a midpoint scheme as discussed below.

The combination of $G_{ex}$ and $G_{im}$ happens through an IMEX scheme. Multiple schemes of different orders are available, for illustrative purposes we show below only the second order accurate Midpoint scheme. This is the scheme that was used for the test cases described in section \ref{sec:Testcases}, unless stated otherwise. The midpoint scheme advances from $\vec{u}^{n}$ to $\vec{u}^{n+1}$ as follows:

\begin{align}
    \vec{u}^{n+1/2}_+ &= \vec{u}^{n} + \frac{1}{2} \Delta t G_{ex}(\vec{u}^{n}) \label{eq:G_expl},\\
    \vec{u}^{n+1/2} &= \vec{u}^{n+1/2}_+ + \frac{1}{2}\Delta t G_{im}(\vec{u}^{n+1/2}), \label{eq:G_impl}\\
    \vec{u}^{n+1} &= \vec{u}^{n} + \Delta t \left( G_{ex}(\vec{u}^{n+1/2}) + G_{im}(\vec{u}^{n+1/2}) \right). \label{eq:G_implexpl}
\end{align}
where $\vec{u}^{n+1/2}$ is the state half a time step later. In the formulation above, \eqref{eq:G_expl} adds the $G_{ex}$ term for half a time step, computed from $\vec{u}^{n}$. Then, half of an implicit time step is added by advancing to $\vec{u}^{n+1/2}$ in \eqref{eq:G_impl}. Finally, in \eqref{eq:G_implexpl}, the full time step is computed by advancing $\vec{u}^{n}$ to $\vec{u}^{n+1}$ where $G_{ex}$ computed from the midpoint state $\vec{u}^{n+1/2}$ and the value of $G_{im}$ is re-used from \eqref{eq:G_impl}. 
Here, in both the first and last operation\,\eqref{eq:G_expl} is performed following the scheme\,\eqref{eq:G_expl_1}-\eqref{eq:G_expl_3} for half a time step and a full time step respectively.

In order to ensure numerical stability, an adapted Courant-Friedrichs-Lewy (CFL) condition  \citep{Courant1928} is implemented. For radiation-free hydrodynamics, a stable time step can be ensured by calculating the local maximum propagation speed of sound waves and limiting the time step in such a way that the distance traveled by a sound wave in one time step is smaller than the width of a cell. In a RHD setting, one has to take into account the propagation of radiative-acoustic waves when computing a constraint on the timestep. To this end, the radiation pressure is added to the gas pressure when calculating a time step constraint.
%
%


\subsection{Radiation force and photon tiring} 
We next outline numerical aspects behind computing and adding the explicitly handled $S_{ex}$ terms, namely radiation force \eqref{eq:rhd_force}, its work term and the photon tiring term. Both the flux limiter $\lambda$ and the radiation flux according to \eqref{eq:fld_F} depend on the gradient of $E$. This gradient is computed using a 4th order central difference with a 5 point stencil \citep{Fornberg88}. E.g., in a 2D set-up, the $x$-component of the gradient reads:
 
  \begin{align}
     \left( \nabla E_{i,j} \right)_x = \frac{\frac{1}{12} E_{i-2,j} - \frac{2}{3} E_{i-1,j} + \frac{2}{3} E_{i+1,j} - \frac{1}{12} E_{i+2,j}}{\Delta x}
     \label{eq:ileyk_flux}.
 \end{align} 
We use equations \eqref{eq:rhd_force} and \eqref{eq:fld_F} to add in an explicit way the source terms for the momentum and gas energy. In this way, the radiative force and its work are not purely local since they are derived from the radiative flux using equation\,\eqref{eq:ileyk_flux} above.
 
Moreover, $\vec{P}:\nabla \vec{v}$, which enters equation \eqref{Eq:0th_cmf}, is calculated by the dyadic product between the radiation energy pressure tensor and the gradient of the velocity vector. Using \eqref{eq: fld_Edd_fac} and \eqref{eq:fld_Edd}, the components of the radiation pressure tensor can be calculated. Here again, the gradients of each of the components of the velocity field are calculated with a similar 5 step stencil for better accuracy. As an example, we provide below the representation of the photon tiring work term in the point $(i,j)$ for a 2D setup: 
 
 \begin{align}
     (\vec{P}:\nabla \Vec{v})_{i,j} &= P^{x,x}_{i,j}  \left( \nabla (v_x)_{i,j} \right)_x + P^{x,y}_{i,j}  \left( \nabla (v_x)_{i,j} \right)_y  \nonumber \\
                              &+ P^{y,y}_{i,j}  \left( \nabla (v_y)_{i,j} \right)_y + P^{y,x}_{i,j}  \left( \nabla (v_y)_{i,j} \right)_x.
 \end{align}
 Also this source term is added explicitly in every cell. For an addition of $\Delta t S_{ex}$:
 
  \begin{equation}
     E^{n+1} = E^n + \Delta t \left( \vec{P}^{n}:\nabla \vec{v}^{n} \right).
 \end{equation}


\subsection{Heating and cooling} \label{subsec:HeatCool}
As mentioned above, our non-equilibrium FLD description allows the gas temperature to be different from the radiation temperature. Locally, the gas temperature can be evaluated from the gas pressure and density using the ideal gas law \eqref{eq:hd_gaslaw}. The radiation temperature is evaluated from $E = a_r T_r^4$. When both temperatures are equal, the system is in radiative equilibrium. However, when the radiation temperature is higher/lower than the gas temperature, the gas will heat up/cool down due to an energy exchange with the radiation field. This energy exchange is written as $\dot{q}$ in equations \eqref{eq:hd_e} and \eqref{Eq:0th_cmf}. The timescale for heating and cooling is typically much shorter than a Courant time step. To ensure numerical stability over a time step,
the gas heating and cooling terms are therefore added implicitly. This approach is based on the method described in \citet{Turner12001}. Since the gas cooling term depends only on the internal gas energy density $\epsilon$ (and is independent of kinetic energy), we first compute the internal gas energy density $\epsilon = e - \rho v^2/2$. Using \eqref{eq:hd_gaslaw}, a point-implicit, discretised formulation of the heating and cooling terms applied to the gas internal and radiation energy density are written as: 

\begin{align}
    \epsilon^{n+1} &= \epsilon^n + \Delta t \left[ c\kappa^n \rho^n E^{n+1} - 4 \kappa^n \rho^n \sigma \left(\frac{m_p \mu}{k_b} (\gamma -1) \frac{\epsilon^{n+1}}{\rho^n}\right)^4  \right],\\
    E^{n+1} &= E^n + \Delta t \left[ -c\kappa^n \rho^n E^{n+1} + 4 \kappa^n \rho^n \sigma \left(\frac{m_p \mu}{k_b} (\gamma -1) \frac{\epsilon^{n+1}}{\rho^n}\right)^4  \right] \label{eq: qdot_gas}.
\end{align}
 The adiabatic index $\gamma$ and mean molecular weight $\mu$ are constant for a given simulation as we do not take into account ionisation effects. 
 Solving these coupled equations comes down to first finding the root of the following $4^{th}$ degree polynomial \citep{Turner12001}:  
 
 \begin{align}
     \left(\epsilon^{n+1} \right)^4 + \left(\frac{1+a_2}{a_1} \right) \epsilon^{n+1} - \left(\frac{1+a_2}{a_1} \right) \epsilon^n - \frac{a_2}{a_1} E^n = 0, 
 \end{align}
 where $a_1 = 4 \kappa \sigma (\gamma - 1)^4 /\rho^3 \Delta t$ and $a_2 = c \kappa \rho \Delta t$, with both $\kappa$ and $\rho$ evaluated at step $n$. To get the gas internal $\epsilon^{n+1}$ energy at the next timestep, Halley's root finding method \citep{NumRec2007} is employed. This method is similar to Newton-Raphson  but uses the second derivative for a faster convergence. If this method does not reach a user set tolerance after a certain number of iterations (typically 100), we switch to a bisection scheme. Equation \eqref{eq: qdot_gas} only has one real positive root which has to be smaller than the sum of the internal and radiation energies and thus lies in the interval $]0, \epsilon^n + E^n[$. With the updated internal gas energy, the radiation energy density is given by:
 
 \begin{align}
     E^{n+1} = \frac{a_1 \left(\epsilon^{n+1}\right)^4 + E^n}{1+a_2}.
 \end{align}

\subsection{Radiation diffusion}
For the diffusion term in equation \eqref{Eq:0th_cmf}, we make use of the newly developed geometric multigrid method library \texttt{octree-mg} \citep{Teunissen2019}. This MPI-parallelised library is capable of solving elliptic PDEs on 1D,2D or 3D Cartesian grids and is fully compatible with the block-tree AMR structure used in \vac. The contribution by the diffusion term can be written in an operator split way by only considering the advection of the CMF radiation flux from equation \eqref{Eq:0th_cmf},  $\partial_t E + \nabla \cdot \vec{F} = 0$,  with $\vec{F}$ substituted using \eqref{eq:ileyk_FLD}: 

\begin{align}
    \partial_t E + \nabla \cdot (-D \nabla E) &= 0.
\end{align}
Due to the elliptic nature of the diffusion term and the difference between the gas and radiation dynamical timescales, it is computationally inefficient to use an explicit solver. Indeed, in many astrophysical contexts, such explicit solvers for radiative diffusion often require prohibitively short timesteps to be stable (typically several orders of magnitude smaller than that for the advection term). As such an implicit method is used here instead. 
Geometric multigrid methods speed up the convergence rate of iterative relaxation by relaxing the PDE solution on a hierarchy of grids. In this way large scale errors can be damped more efficiently on a coarser grid, while small scale errors are smoothed on a finer grid. The multigrid library uses a standard Gauss-Seidel smoother over multiple grid levels to solve a Helmholtz equation.

The diffusion problem can be recast as a Helmholtz equation with a variable diffusion coefficient, where $E^{n}$ is the current radiation energy density and $E^{n+1}$ is the radiation energy density in the next time step: 

\begin{align}
    E^{n+1}/\Delta t - \nabla \left( D^n \nabla E^{n+1} \right) = E^n/\Delta t.
\end{align}
As mentioned above, our methods are implemented in 1D, 2D and 3D. In 2D, this equation is discretised using the following 5 point stencil: 
 
 \begin{align}
     \left[ D^n_{i-1/2,j} (E^{n+1}_{i,j} - E^{n+1}_{i-1,j}) - D^n_{i+1/2,j} (E^{n+1}_{i+1,j} - E^{n+1}_{i,j}) \right] / \Delta x^2 \nonumber \\
     + \left[ D^n_{i,j-1/2} (E^{n+1}_{i,j} - E^{n+1}_{i,j-1}) - D^n_{i,j+1/2} (E^{n+1}_{i,j+1} - E^{n+1}_{i,j}) \right] / \Delta y^2 \nonumber \\
     + E^{n+1}_{i,j}/\Delta t =  E^{n}_{i,j}/\Delta t. \label{eq: discdiff}
 \end{align}
In addition to an initial state $E^n$, in order to advance the radiation energy density, one has to specify boundary conditions at every boundary. This can be done by either defining the value $E^{n+1}$ in the ghost cells (Dirichlet conditions), by assuming a fixed gradient of $E^{n+1}$ at the interface between the ghost cells and the numerical domain (Neumann conditions), or by a linear extrapolation into the ghost cells (continuous boundary conditions).
 
 In \eqref{eq: discdiff}, the diffusion coefficient is evaluated between two neighbouring cells using a harmonic mean as described by \citet{Teunissen2019}: 
 
 \begin{equation}
     D^n_{i-1/2,j} = \frac{2 D^n_{i,j} D^n_{i-1,j}}{D^n_{i,j} + D^n_{i-1,j}}. 
 \end{equation}
 Locally, the diffusion constant in the cell centers is calculated from equation \eqref{eq:fld_F}, where the flux-limiter $\lambda$ is computed according to \eqref{eq:fld_lambda}. 
 First, the diffusion solver will try a full multigrid (FMG) cycle to converge to a requested residual which for the tests described below was typically chosen to be $10^{-5}$. If the FMG-cycle does not reach this tolerance, additional V-cycles are added until the convergence criterion is met.
 
Though fully compatible with the adaptive mesh refinement, this FLD module is currently limited to Cartesian settings due to the abilities of the multigrid solver used to solve the diffusion part of equation \eqref{Eq:0th_cmf}. In the future, we plan to expand this module to the cylindrical, polar and spherical meshes which are already available for the usual non-radiative HD/MHD setup. These limitations in the multigrid solver arise from using point-wise relaxation methods in smoothing the error in the solution. These relaxation methods are not efficient when used on spherical diffusion operators, as discussed in \citet{Teunissen2019} and \citet{Briggs00}.
 

\section{Results: Testcases} \label{sec:Testcases}

In this section we test aspects of the newly developed FLD code in various regimes. Since non-trivial analytic solutions to the FLD-equations are scarce, we set up a number of test-cases and then compare to various semi-analytic solutions or predictions (see also, for example, \citealt{Turner12001, Krumholz2006}). In section \ref{subsec:H&C} the implicit heating and cooling terms are tested, for which the algorithm has been explained in section \ref{subsec:HeatCool}. 
Both tests in sections \ref{subsec:RHDshock} and \ref{subsec:RHDwave} test the full system of equations, in a steady state and dynamic state respectively. In section \ref{subsec:GalInv} we will focus on a test covering the Galilean invariance of the code. 
Finally, the test described in section \ref{subsec:OpThin} examines the optically thin limit.

\subsection{Heating and Cooling} \label{subsec:H&C}
As a first test, we check the energy exchange by means of heating and cooling in a gas with zero velocity and a constant density. The 2D domain has periodic boundary conditions on all sides. The gas energy and radiation energy are initialised out of radiative equilibrium, so the gas temperature is not equal to the radiation temperature. Due to this non-equilibrium, there is a net heating/cooling term $\dot{q}$, which will relax the gas energy density to its equilibrium value and rise/lower the gas and radiation temperatures until they equal. Since there is no velocity, there is no kinetic energy and the internal gas energy is equal to the total gas energy $\epsilon = e$.

We consider a gas with density $\rho = 10^{-7} \, \text{g} \, \text{cm}^{-3}$, opacity $\kappa = 0.4 \, \text{cm}^2 \, \text{g}^{-1}$, adiabatic index $\gamma = 5/3$ and mean molecular weight $\mu = 0.6$. The radiation energy density is initialised as $E = 10^{12} \, \text{erg} \, \text{cm}^{-3}$.

\begin{figure}
\centering
\includegraphics[width=\hsize]{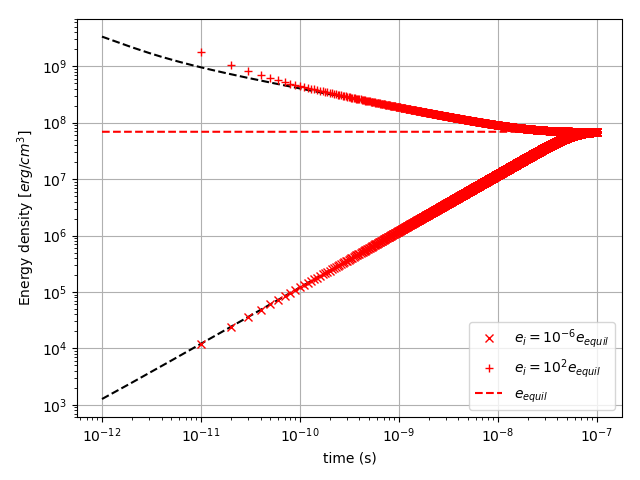}
  \caption{
  This figure shows, as function of time, the gas energy densities of a gas that is relaxing to radiative equilibrium from two sets of initial conditions. Marked with a '+', relaxation where the initial gas energy is set as $10^2$ times the expected equilibrium value. Marked with a 'x', the initial gas energy is set as $10^{-6}$ times the equilibrium value. Since the time axis is logarithmic, the initial conditions at $t=0$ are not visible.
  }
     \label{fig:heatcool}
\end{figure}

The equilibrium energy density $e_{equil}$ for gas and radiation can be calculated from conservation of the sum of the initial energies and the condition that there is no net cooling or heating when the system is in equilibrium. We consider two different uniform initial conditions for the gas energy density relative to this equilibrium energy density. In the first test, $e$ is set to $10^{-6} e_{equil}$ ('x' symbols in figure\,\ref{fig:heatcool}) and in the second test, it is set to $10^2 e_{equil}$ ('+' symbols). Thereafter, we let the system relax to equilibrium and monitor the gas and radiation energy through time.\\

In figure \ref{fig:heatcool}, the gas energy density is plotted through time on a log-log plot, together with the energy equilibrium value. For this test, it is important that the rate at which the temperature approaches its equilibrium is correct. To this end, we compare the previous simulation for two different fixed time steps: once where $\Delta t = 10^{-12} \, \text{s}$ which is regarded as a physically correct baseline, and once where $\Delta t = 10^{-11}\, \text{s}$. From the resulting curves, we can conclude that in the test where the gas energy density started out lower than the equilibrium energy and that it relaxes to the theoretically predicted value $e_{equil}$ on a timescale which is independent of the used numerical time step. However, for the case where the gas energy is initiated greater than the equilibrium value, the heating lags behind for a shorter time step in the first couple of iterations. Later on, the correct equilibrium value is found on a correct timescale. Note that the time step of $10^{-11} \, \text{s}$ used here is greater than the initial cooling timescale $\tau_{cool} \sim e/(4\pi\kappa \rho B) \sim 10^{-12} \, \text{s}$. 

Although the radiative energy density is left free to evolve, it remains essentially constant because its initialized value is much larger than the gas energy density. However, we verified that it matches with the equilibrium value for radiative energy.

\subsection{Radiation dominated shock} \label{subsec:RHDshock} 
This test case here describes a steady state radiation dominated shock. Although there is no analytical solution for the exact shape of the shock front, there are approximations for its expected width \citep{Mihalas1984a}. Using radiation-modified Rankine-Hugoniot conditions, a left and right hand side gas state are calculated and used as initial condition, before they are relaxed to a steady state. With this test, we can asses how the code handles conserved quantities. Moreover, the discontinuity is also an ideal situation to test the adaptive mesh refinement (AMR).

\begin{figure}
  \centering
  \includegraphics[width=\hsize]{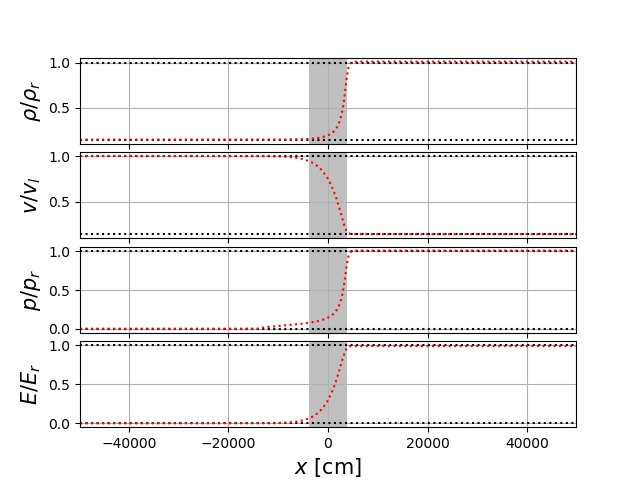}
      \caption{Density, velocity, gas pressure and radiation energy profiles for a 1D simulation of a radiation dominated shock after 10 flow passing times. The profiles are normalised to the initial right hand side values, except for the velocity profile, which has been normalised to the initial left hand side value. The expected width of the shock is illustrated in grey.}
         \label{fig: rhd_shock}
  \end{figure}

On the left hand side of the shock, gas density $\rho_l = 10^{-2} g \, \text{cm}^{-3}$, velocity $v_l = 10^{9} \text{cm} \, \text{s}^{-1}$ and total (kinetic + internal) gas energy density $e_l = 5.0 \, 10^{15} \, \text{erg} \, \text{cm}^{-3}$. The radiation energy density is set in equilibrium to $E_l = 75.6 \, \text{erg}  \, \text{cm}^{-1}$. On the right hand side $\rho_r = 6.85874 \, 10^{-2} \, \text{g} \, \text{cm}^{-3}$, $v_r = 1.45 \, 10^8 \, \text{cm} \, \text{s}^{-1} $, $e_r = 1.93 \, 10^{15} \, \text{erg} \, \text{cm}^{-3}$ and $E_r = 2.44 \, 10^{16} \, \text{erg} \, \text{cm}^{-3}$. The stability of the shock depends on having the initial values conform with the radiation-modified Rankine Hugoniot conditions. Empirically, at least five decimals for the right hand side density were needed to converge to these profiles. We assume we are in the optically thick diffusion limit so that $\lambda = 1/3$. Both the radiation flux and radiation force will point upstream, widening the shock to a width which has been predicted to be $w = D/v_l$ \citep{Mihalas1984a, Turner12001}. We consider a 
fully ionised pure hydrogen gas 
with $\mu = 0.5$ and $\gamma = 7/5$. The simulation box is $10^5 \, \text{cm}$ wide and consists of 256 cells on the lowest AMR level (level 1). In figure \ref{fig: rhd_shock}, we show the relaxed shock after 10 flow passing times. 

As can be seen in figure \ref{fig: rhd_shock}, the shock width of the 1D simulation is conform with the predictions made by \citet{Mihalas1984a}. We obtained the same results for both 2D and 3D simulations of the shock along the $x$-axis, which have been performed but are not shown here. Moreover, one can check whether the solution is truly steady by comparing the momentum on both sides of the shock. From this, we retrieve a $0.2 \%$ relative error for a low resolution run on $N_x = 64$ without any AMR, a $0.1 \%$ relative error for a high resolution run on $N_x = 256$ without any AMR and similarly a $0.1 \%$ relative error for a run with base resolution of $N_x = 64$ but an effective resolution of $N_x = 256$ by using $3$ levels of refinement.

\subsection{Galilean invariance} \label{subsec:GalInv}
\begin{figure}
  \centering
  \includegraphics[width=\hsize]{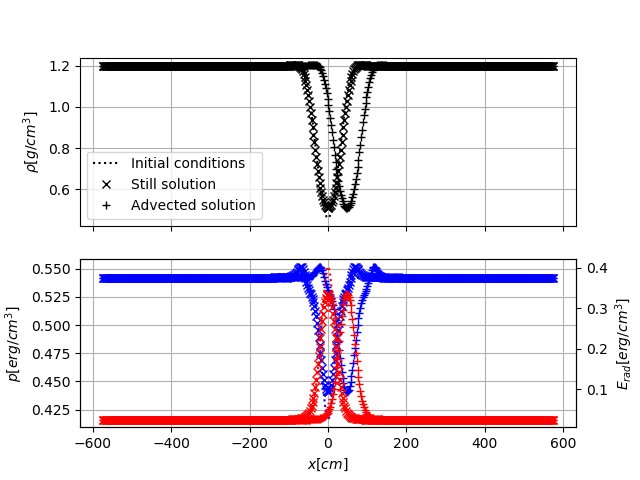}
      \caption{In the top panel density profiles corresponding to the Galilean invariance test. Initial conditions (in dashed lines) are relaxed once with (marked with '+') and once without (marked with 'x') an initial background velocity field. In the bottom panel we see the same but for gas pressure (blue) and radiation energy density (red). The horizontal shift between the curves correspond to the expected rightward displacement.}
         \label{fig: rhd_pulse}
  \end{figure}

In this test we check the Galilean invariance of the code during an advection diffusion problem. Again, there is no available analytic solution, but if Galilean invariance is respected, the two profiles should keep the same shape. The same simulation is performed twice: once with and once without an initially constant background velocity field. If solved for correctly, the profiles of the conserved quantities in the two simulations will be the same but translated. Following \citet{Krumholz2006}, we consider a slab of gas in total pressure equilibrium. In the center of the gas, there is a dip in density and gas pressure, and a corresponding bump in radiation energy density and pressure to preserve a constant total pressure. When the simulation is started, the radiation will diffuse out of the dip, the total pressure equilibrium is lost and gas will be driven towards the center, where the gas pressure is lower. The initial conditions are given by:

\begin{align}
    T   &= T_0 + (T_1 - T_0) \exp\left(\frac{-x^2}{2w^2}\right), \\
   \rho &= \rho_0 \frac{T_0}{T} + \frac{a_r \mu}{3k_b}\left(\frac{T_0^4}{T} -T^3\right),
\end{align}
where $T_0 = 10^7 \text{K}$, $T_1 = 2 \, 10^7 \text{K}$, $\rho_0 = 1.2 \, \text{g} \, \text{cm}^{-3}$ and $w = 24 \, \text{cm}$. Both $E$ and $e$ are set in equilibrium to the above temperature profile. The mean molecular weight is taken to be $\mu = 2.33$, which corresponds to a Helium abundance of $0.1$. The adiabatic exponent $\gamma = 5/3$ and $\kappa = 100 \, \text{cm}^2 \, \text{g}^{-1}$.
This simulation is ran twice: a first time with zero background velocity and a second time with a constant background velocity of $v=5 \, 10^7 \, \text{cm} \, \text{s}^{-1}$. In figure \ref{fig: rhd_pulse}, we show both solutions after approximately 1 pulse width crossing times. 

    \begin{figure}
  \centering
  \includegraphics[width=\hsize]{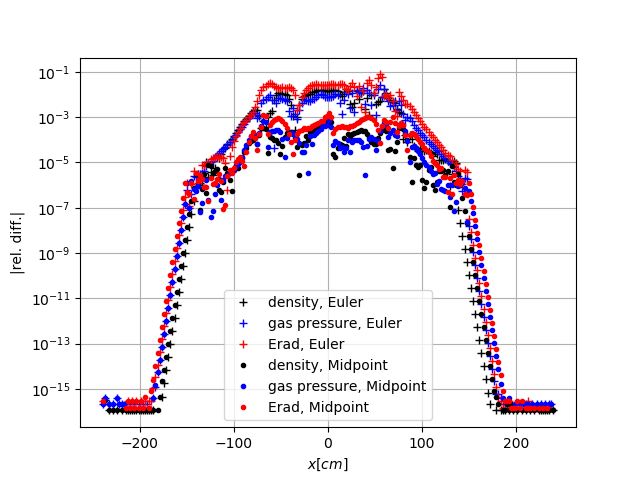}
      \caption{Absolute value of the relative difference between the profiles with and the profiles without an initial background velocity field from the Galilean invariance test in figure \ref{fig: rhd_pulse}, where the advected profile is translated back to the origin. We compare two different time integration schemes: A first order Euler IMEX scheme (marked '+') and a second order Midpoint IMEX scheme (marked '$\cdot$'). Different colors correspond to different quantities: gas density in black, gas pressure in blue and radiation energy in red.}
         \label{fig: rhd_pulse2}
  \end{figure}
  
As can be seen from figure \ref{fig: rhd_pulse2}, the relative difference between the advected and stagnant density profile peaks at $0.03\%$ when using a second order Midpoint IMEX-scheme. However, when using a simpler first order Euler scheme, the relative difference in density goes up to $1.7\%$. Similar improvements can be seen for the relative differences in gas pressure and radiation energy density. This clearly shows the advantage of using higher order time stepping schemes.

\subsection{Optically thin limit} \label{subsec:OpThin}
While all of the previous tests were situated in an optically thick regime, the following assesses the workings of the FLD code in an optically thin situation. This is an important test, since if not flux-limited, the diffusion approximation breaks down in this regime.

Due to the flux limiter $\lambda$ we should recover the free streaming flux $|\textbf{F}| = cE$ when the optical depth approaches zero (ensuring the radiation energy density does not travel faster than the speed of light). In this test case we start with a slab of gas with $\rho_0 = 0.025 \, \text{g} \, \text{cm}^{-3} $ and constant opacity $\kappa = 0.4 \, \text{cm}^2 \, \text{g}^{-1}$. The numerical domain reaches from $x= -0.5 \, \text{cm}$ to $x= 1.5 \, \text{cm}$ and it is subdivided in $N_x = 256$ grid cells. This gives a total optical thickness of the slab of $\tau = 0.002 $. The initial radiation energy density is set according to an error function, centered around $x=0$ with a width $d= 0.05 \, \text{cm}$:

\begin{align}
    E(x) &= E_0 + \frac{1}{2}\left[1-erf\left(\frac{x}{d}\right)\right]E_1. 
\end{align}
This expression lights up the slab from the left hand side, with $E_1 = 1.4 \, 10^{11} \, \text{erg} \, \text{cm}^{-3}$. There is no velocity and the gas energy everywhere is set to be in equilibrium with $E_0 = 1.4 \, 10^{-11} \, \text{erg} \, \text{cm}^{-3}$. For this setup, we only perform the radiation diffusion (thus the radiation force, heating, cooling and photon tiring terms are not considered). Since the local sound speed is not relevant for the free streaming radiation field, the time step is set manually to $\Delta t = 10^{-13} \, \text{s}$, which is on the order of half a cell-crossing time for the speed of light. This time step is very stringent but is chosen only for the sake of illustrative purposes. In practice, larger time steps can be used. 

  
\begin{figure}
\centering
\includegraphics[width=\hsize]{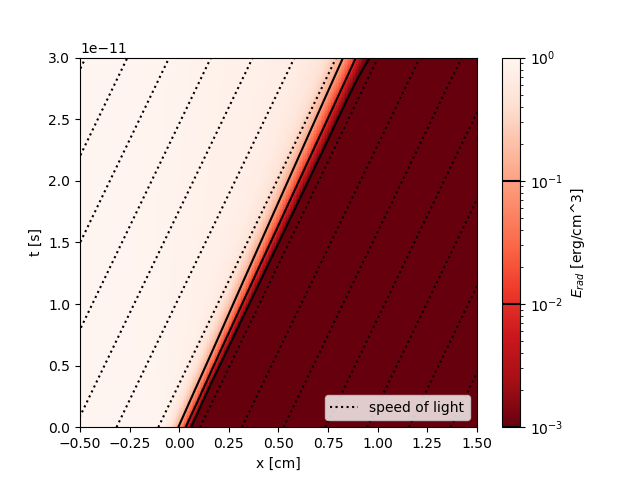}
    \caption{A space-time diagram of the radiation energy front evolving in an optically thin slab of gas. In the optically thin limit, the FLD flux limiter $\lambda$ prohibits radiation from moving faster than the speed of light. In this plot, the speed of light has been indicated by the black dotted lines. In addition to the colormap, the position of the radiation front has been evaluated at three arbitrary values with the red contour lines. The total optical thickness of the slab is 0.002.}
         \label{Fig: propfront_xt}
\end{figure}

 The propagation speed of the radiation energy density is, as expected, limited by the speed of light as can be seen in figure \ref{Fig: propfront_xt}. The initial profile has diffused a little, which is to be expected when executing what is practically an advection operator with a diffusion solver.
  
\subsection{Linear RHD wave} \label{subsec:RHDwave}
When interacting with radiation, acoustic waves can be naturally damped radiating away energy. This process can be modelled with the FLD description of RHD. With the following RHD wave setup, the full system of equations can be tested against results of a semi-analytic perturbation relation. In addition to checking the different algorithms for adding the different source terms described in section \ref{sec:NumImpl}, this setup allows us to assess the numerical diffusion of the code. We also compare the 1D problem with a 2D setup which is symmetric along the first diagonal of the grid, which allows us to rule out any effects of grid anisotropy on the simulation outcome.

We test the code with dispersion relations that come out of an analytical linear perturbation analysis of the RHD system in the diffusion limit done by \cite{Mihalas1984a}. This analysis shows that the damping length of a running wave is dependent on two dimensionless parameters. The Boltzman number $Bo \sim 4\gamma c_a e / (cE)$, with $c_a = \sqrt{\gamma p / \rho}$ the adiabatic sound speed, translates to how much radiation contributes to heat transfer. The ratio of energy densities $r \sim E/(4 \gamma e)$ is large when a gas is radiation energy dominated and low when gas is gas energy dominated. 

A slab of gas is considered with background values $\rho_0 = 3.216 \, 10^{-9} \, \text{g} \, \text{cm}^{-3}$, $e_0 = 26.02 \, 10^3 \, \text{erg} \, \text{cm}^{-3}$ and $E_0 = 17.34 \, 10^3 \, \text{erg} \, \text{cm}^{-3}$. The adiabatic index is $\gamma = 5/3$ leading to a Boltzman number of $Bo = 10^{-3}$ and energy density ratio $r = 0.1$.  The slab is perturbed on the left hand side with the following boundary conditions to excite a traveling wave:

\begin{align}
    \rho &= \rho_0 + A_\rho \sin(k x - \omega t), \\
    \textbf{v} &= A_v \sin(k x - \omega t), \\
    e &= e_0 + A_e \sin(k x - \omega t). 
\end{align}
Here, the wave number $k$ is chosen in such a way that for the constant opacity of $\kappa = 0.4 \, \text{cm}^2 \, \text{g}^{-1}$, the optical depth across one wavelength is $\tau_\lambda = 10^3$. The pulsation $\omega$ is related to $k$ via the propagation speed of the RHD wave. The perturbation with density amplitude $A_\rho = 0.01 \, \rho_0$ will travel along through the gas, but it will be damped by the effects of the radiation field. Amplitudes for the velocity and gas energy perturbation are set by $A_v = A_\rho \omega/(k \rho_0) $ and $A_e = A_\rho e_0/\rho_0$. Gas energy in the oscillation will be transferred to the radiation field by means of the cooling mechanism. After this, the radiation energy will leak outward and leave the system due to diffusion. The dispersion relation given by \cite{Mihalas1984a} can be solved with a standard root-finding algorithm to provide a theoretical dampening length for the oscillation. 
For the numbers provided here, the predicted dampening length is 8.16 $\lambda$, with $\lambda$ the wavelength of the induced acoustic oscillation (not to be confused with the flux-limiter introduced in section\,\ref{sec:ileyk_FLD}).


  \begin{figure}
  \centering
  \includegraphics[width=\hsize]{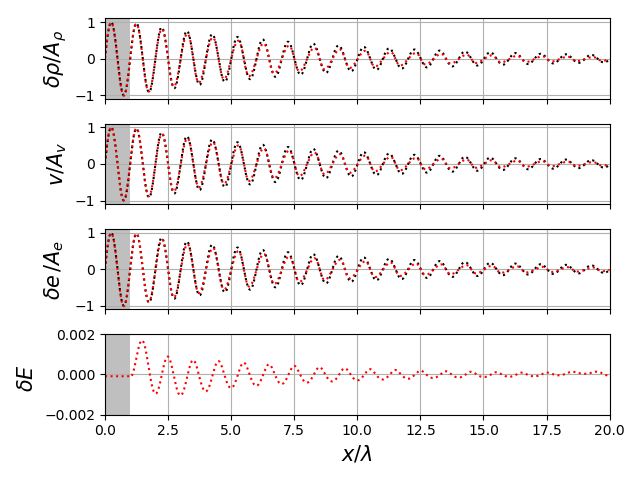}
      \caption{From top to bottom: the density perturbation, velocity, gas energy perturbation and radiation energy perturbation after 40 oscillation periods, for a RHD wave where the optical thickness of a wavelength is equal to 1000. For the top three panels this is compared to the analytic profile (black dotted lines) calculated as described by \citet{Mihalas1984a}. The wave is driven from the grey shaded area on the left side of the plot.}
         \label{fig: rhd_wave1}
  \end{figure}
  
    \begin{figure}
  \centering
  \includegraphics[width=\hsize]{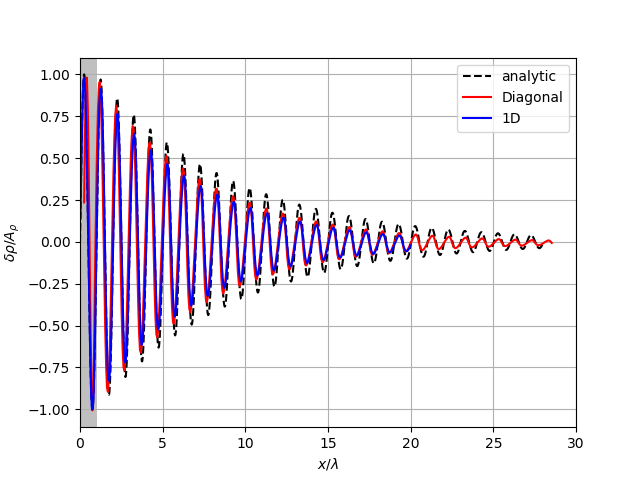}
      \caption{Density perturbation profile of a RHD wave after 40 oscillation periods. In black, the analytically predicted profile, in blue results from a 1D simulation and in red results from a 2D simulation with the wave vector parallel to the first diagonal. The $x$-axis is here chosen in the direction of the wave vector.}
         \label{fig: rhd_wave2}
  \end{figure}
  
    \begin{figure}
  \centering
  \includegraphics[width=\hsize]{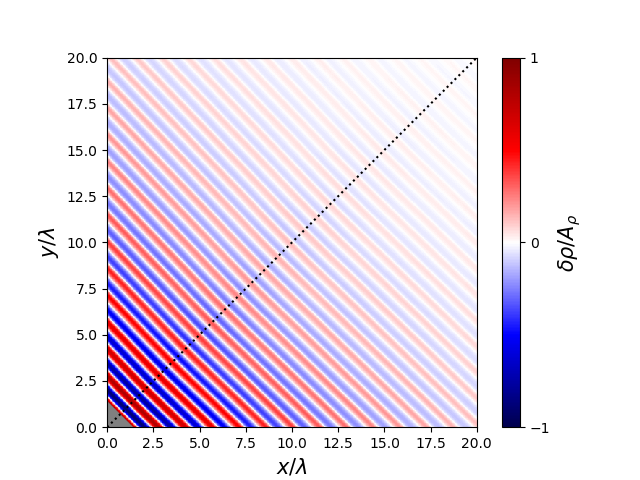}
      \caption{A map of density perturbation in a RHD wave along the first diagonal (black dotted line) of a 2D plane, after 40 oscillation periods. The wave is driven from the grey area in the bottom left corner.}
         \label{fig: rhd_wave3}
  \end{figure}
  
 In figure \ref{fig: rhd_wave1}, results are shown for a perturbation with the optical depth  $\tau_\lambda = 10^3$ in a 1D setup. For the 2D version, the same problem is set up along the line of the first diagonal. Now, the wave is driven at any point where $x + y < \lambda$. Instead of $(k x - \omega t)$, the sine in the driving conditions now take  $(k (x+y) /\sqrt{2} - \omega t)$ as an argument. Additionally, boundary conditions are copied from their diagonal neighbours for all conserved quantities, such that they correspond to the correct phase in the diagonally traversing wave. For the left and top boundary, the conserved quantities in cell $(i,j)$ are copied from their bottom right neighbour: $\vec{u}_{i,j} = \vec{u}_{i+1,j-1}$. For the right and bottom boundary, they are copied from the top left neighbour: $\vec{u}_{i,j} = \vec{u}_{i-1,j+1}$.
  
The solution depends heavily on the flux limiter that is used in the approximate Riemann solver. More diffusive schemes such as minmod \citep{Roe86} or Koren \citep{Koren1993} give a shorter dampening length, while more advanced, higher order schemes such as a 5th order weighted essentially non-oscillating scheme \citep{Liu94} or monotonicity preserving \citep{Suresh97} are better at approaching the correct signal. Overall, figures \ref{fig: rhd_wave1} and \ref{fig: rhd_wave2} show that the numerical solution is a very good match to the first order perturbation relation, both in wavelength and dampening length. As seen from figure \ref{fig: rhd_wave3}, the solution stays symmetric along the first diagonal, as expected.

\section{First research application: Wolf Rayet wind} \label{sec:StellarAppl}

As a final display of our code's  applicability, we perform a 1D simulation of a supersonic, optically thick Wolf-Rayet wind outflow. Classical Wolf-Rayet stars are massive stars that have evolved back to the blue side of the Hertzsprung-Russel diagram, after shedding their outer hydrogen layers thus exposing a helium core (see review by \citealt{Crowther2007}). Observationally they are known to have supersonic wind outflows characterised by high mass-loss rates. Wolf-Rayet winds are believed to be accelerated by radiation \citep{Sander2020,Poniatowski2020}, but due to the high mass-loss rate their hydrostatic surface lies deep within optically thick layers. This provides an interesting first research-application for the FLD code, testing the effects of a strong radiation force in an optically thick and highly supersonic environment.

The simulation set-up is based on the recent work by \citet{Poniatowski2020}, but now using the FLD method described in previous sections to compute the radiation force and energy balance. This then allows for a more flexible (and complete) approach for time-dependent modeling of such Wolf-Rayet outflows. Namely, while the simulations by \citet{Poniatowski2020} assumed that
the local radial flux always was set by $L_\ast/(4 \pi r^2)$, where $L_\ast$ is the stellar core luminosity, in the FLD method presented here fluxes (and radiation work terms, neglected in Poniatowski et al.) are computed directly from the evolving energy density. As such, in contrast to the Poniatowski et al. model, the FLD method presented here could be readily extended to time-dependent 2D or 3D flows with local (and potentially non-radial) fluxes and forces, as further discussed below. Also, a spherically symmetric wind is assumed, hence the Cartesian formulation of the RHD equations has to be corrected for spherical geometry. For this, we follow the recipe by \citet{Sundqvist2018}, as further explained in the appendix.    

As in \citet{Poniatowski2020}, for this 
first 1D Wolf-Rayet simulation we assume a fixed stellar mass $M_* = 10 \, M_\odot$, a gravity source term $f_g = \rho G M_\ast/r^2$, a hydrostatic core lower boundary radius $R_* = 1 \, R_\odot $, and a stellar core luminosity $ log_{10} (L_*/L_\odot) = 5.416$.  


\subsection{Opacities}
Since the WR outflow is initiated by the radiation force, a key feature in this model regards the applied opacities. To this end, we follow \citet{Poniatowski2020} and use a superposition of equilibrium opacities computed in the static limit and a simple parametrised form for the large enhancement of line-opacity expected in a supersonic flow:   

\begin{align}
    \kappa = \kappa^{OPAL}\left(\rho, T_g\right) + \kappa^{CAK}\left(\rho,dv/dr\right). 
\end{align}
Here $\kappa^{OPAL}$, which represents opacities computed for static media, is taken from the tabulations by \citet{Iglesias1996} and $\kappa^{CAK}$ uses a variant of the parametrisation first introduced by \citet{CAK1975} ('CAK') to represent the accumulative effect of Doppler shifted lines. At every time step, the CAK-opacity is computed locally using a second order central difference derivative of the radial velocity with respect to radius \citep{udDoula02}:

\begin{align}
    \kappa^{CAK}_i = \frac{\kappa_e \bar{Q}}{1-\alpha(r)} \left( \frac{1}{c \kappa_e \bar{Q} \rho_i} \left|\frac{v_{i-1} - v_{i+1}}{r_{i-1} - r_{i+1}} \right|\right)^{\alpha(r)}.
    \label{Eq:kap_cak} 
\end{align}
This combined opacity is then used in all source terms (radiative force, heating/cooling, photon-tiring) as well as in the computation of the diffusion coefficient.

In equation \eqref{Eq:kap_cak}, $\kappa_e \bar{Q}$ represents the line opacity in the limit that all lines would be optically thin, and $\alpha$ represents the slope of the underlying assumed power-law distribution of lines. In general, these parameters should be derived from excitation and ionisation calculations using full line lists \citep{Puls2000, Lattimer2021}. In this first application, however, we assume the same set of parameters as in \citet{Poniatowski2020}, which means we also here have introduced a radial variation of $\alpha$. In the inner wind, for $0< 1- R_*/r < 0.4$, the exponent is set constant at $\alpha = 0.66$. In the outer wind, for $0.65 < 1- R_*/r < 1$, $\alpha = 0.5$, to ensure a steady outflow. In the transition region, where $0.4 < 1- R_*/r < 0.65$, $\alpha$ decreases linearly as a function of $1- R_*/r$ from $0.66$ to $0.5$.

As discussed in that paper, this assumed variation makes the radial outflow stable against fallback by ensuring an outer-wind radiation force strong enough to accelerate the gas towards infinity. 





\subsection{Initial and boundary conditions}
The initial conditions for the density and radial velocity in this simulation are derived from a  constant mass loss rate $\dot{M} = 4 \pi r^2 \rho v_r$ and a so-called $\beta$-velocity law $v(r)$ (see below). From $L_\ast = 4 \pi r^2 F_r$ and the Eddington approximation $F_r = \frac{c}{3 \kappa_0 \rho} \nabla E$ with a constant opacity (set to the electron scattering value for a fully ionised helium plasma), we obtain initial conditions for $E$ by integrating the radiation energy density inward from the outermost point of the simulation where we assume a known floor radiation temperature $T(r_{max}) = 5 \times 10^4 \, \rm K$. Finally, the ideal gas law is used to compute the gas pressure everywhere by assuming thermal equilibrium with the radiation. Putting this together, we obtain for our initial conditions: 

\begin{align}
    \rho(r) &= \frac{\dot{M}}{4 \pi r^2 v}, \\
    v(r) &= v_\infty \left(1-\frac{R_\ast}{r}\right)^\beta, \\
    E(r) &= a_r \left(T(r_{max})\right)^4 + \int_{r_{max}}^r \frac{3 \kappa_0 \rho L_*}{4 \pi r^2 c} dr, \\
    p(r) &= \frac{k_b \rho(r)}{m_p \mu}  \left(\frac{E(r)}{a_r}\right)^{\frac{1}{4}},  
\end{align}
where we set $v_\infty = 1000$ km/s and $\beta=1/2$ for the initial velocity field. 
We note, however, that it is clear that the relaxed steady state will not resemble this simple $\beta$-law model; the conditions above just provide a good setup for initiating the actual simulation.

At the lower boundary, which is subsonic and so bound to the star, we follow the basic setup of previous radiation-driven wind simulations and fix the gas density while letting the velocity float \citep{Sundqvist2013, Driessen2019, Poniatowski2020}. However, here we now additionally need to set the boundary condition for the radiation energy density $E$. 
Instead we use the fixed stellar luminosity to obtain the gradient of the radiation energy, using the diffusion coefficient calculated at the previous time step. A simple finite difference then gives for the lower boundary energy density:  

\begin{align}
    E^n_{i-1} = E^n_i + \frac{L_*}{4 \pi r_i^2 D^{n-1}_i} \Delta r. 
\end{align}
%
Finally, the gas energy in the ghost cells is set to be in equilibrium with the radiation field.

At the supersonic outer boundary, density, momentum and gas energy are extrapolated. The radiation energy density is set here by first computing the local optical depth and from this the radiation temperature, with the optical depth obtained by analytic inward integration from $r = \infty$ assuming the wind has reached its asymptotic velocity at the outer boundary and 
again that the opacity outside this 
is a constant set by electron scattering. 

The simulation is ran on a 1D Cartesian grid which stretches from $1R_*$ to $11R_*$. The FLD solver module is only constructed for such Cartesian geometry. A 1D stellar outflow, however, is a spherically symmetric problem. For this reason, a correction term for the spherical divergence is added to all conservation equations following \citet{Sundqvist2018}, as explained further in the appendix. To increase the resolution near the base (important to resolve the subsonic region), a constant refinement is used in the first 2 stellar radii. On the coarsest level, the $10R_*$ are resolved by $2000$ grid cells. For this simulation we use a 3-step scheme with a TVDLF solver and a minmod slope limiter. 

\subsection{Relaxed profiles}

\begin{figure}
\centering
\includegraphics[width=\hsize]{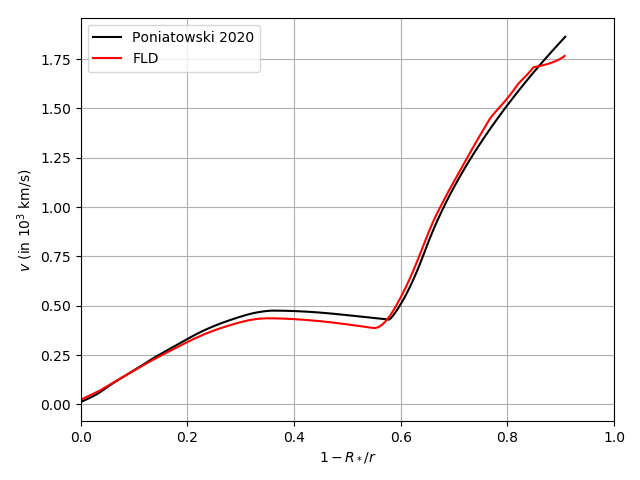}
  \caption{
  Radial velocity profile of a spherically symmetric Wolf-Rayet outflow in a relaxed state. In red, the profile is computed with the 1D FLD method while in black, it is computed following the alternative method by \citet{Poniatowski2020}. }
     \label{fig: WR_1D_v}
\end{figure}

The above described simulations are run until they reach a relaxed steady state.
Figure \ref{fig: WR_1D_v} shows a resulting relaxed velocity profile calculated with our new FLD module, comparing this with the velocity profile calculated by \citet{Poniatowski2020}. The main feature, the stagnated and non-monotonic velocity profile, is similar between the two methods. Both profiles reach the same terminal velocity, however the bump in the FLD profile has a slightly lower velocity than that from \citet{Poniatowski2020}. The resulting stable mass-loss rate for the FLD model is $1.87 \, 10^{-5} M_\odot yr^{-1}$, which again is very similar to the $1.64 \, 10^{-5} M_\odot yr^{-1}$ found by \citet{Poniatowski2020}.

\begin{figure}
\centering
\includegraphics[width=\hsize]{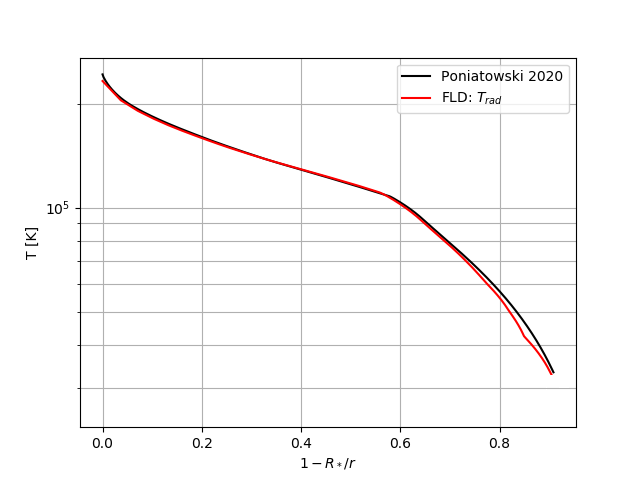}
  \caption{
  Temperature profile of a spherically symmetric Wolf-Rayet outflow in a relaxed state. In red, the profile is computed with the 1D FLD method and in black, it is computed following the alternative method by \citet{Poniatowski2020}. 
  }
     \label{fig: WR_1D_T}
\end{figure}

\begin{figure}
\centering
\includegraphics[width=\hsize]{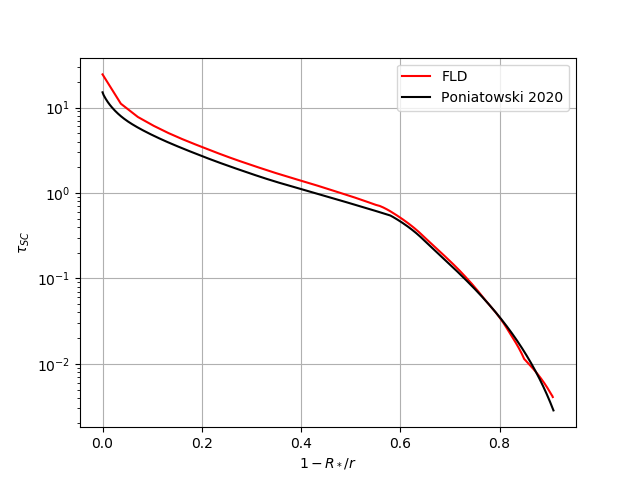}
  \caption{   
  Spherically corrected optical depth of a spherically symmetric Wolf-Rayet outflow in a relaxed state, calculated using the total opacity. In red, the profile is plotted as performed with the 1D FLD method and in black as computed by \citet{Poniatowski2020}. In the latter, this is used to calculate the temperature profile.}  
     \label{fig: WR_1D_tau}
\end{figure}

Figure \ref{fig: WR_1D_T} shows a similar comparison, but for the radiation temperature structure, where qualitatively the models match at the boundary density, in the region experiencing "wind blanketing" from the additional CAK force, and in the outer wind. Finally, in figure \ref{fig: WR_1D_tau}, the optical depth through the stellar wind is computed. Here again, the spherically corrected total opacity was used (see eqn. 8 in \citealt{Poniatowski2020}). As seen from this figure, the simulation spans a wide domain, from the thick core at an optical depth $\tau_{sc} \sim 20$ to the optically thin outer wind where $\tau_{sc} \sim 0.005$. \\

In the deepest layers of the simulation, near the hydrostatic core, acceleration is essentially ensured by the OPAL opacity while in the region transiting towards the optical photosphere and beyond, the CAK opacity dominates. It highlights the complementary role played by the components: while the OPAL opacity lifts up the material from the dense and hot inner regions, resonant line-absorption not only prevents the flow from falling back but also provides it with additional momentum. When the gas reaches the photosphere, the outflow is already highly supersonic. Overall, the wind launching and final escape is thus made possible thanks to the joint action of both opacities. 

In the interest of understanding radiation-powered outflows by means of time-dependent RHD modeling, this simulation illustrates the need for treating the enhanced line-opacity effect in supersonic flows. It also opens the door to the study of time-variable configurations. For instance, it is likely that, when run in a multi-dimensional set up, the lateral symmetry of these Wolf-Rayet models will be broken (see also discussion in Poniatowski et al. 2021). As discussed in the next section, this will then allow us to study structure formation in a radiation-dominated supersonic environment. 



\section{Summary and perspectives} \label{sec:Conclusions}

In this paper, we described the implementation of a radiation module for the finite volume magneto-hydrodynamics code \texttt{MPI-AMRVAC}. We validated it with a set of classic benchmark tests and applied it to a more realistic setup, wind launching and mass loss in the supersonic, expanding atmospheres of Wolf-Rayet stars. The coupling between matter and radiation is performed in the diffusion approximation which provides a closure relation that enables us to deduce the radiative flux and radiative pressure from the energy density of the radiation field. Flux-limiting is applied in order to retrieve the free streaming limit in the optically thin regime, while smoothly transiting to a fully diffusive behavior in highly optically thick environments. The time-dependent evolution equation for the radiative energy density is solved in the co-moving frame to alleviate the angle-dependence of emission, absorption and scattering induced by the Doppler effect. By default, local thermodynamical equilibrium is not assumed and heat exchanges between matter and the radiative field are accounted for. The opacities which enter the formalism (i.e, the energy, Planck, and flux means) can be prescribed a priori or dynamically computed based on hydrodynamical quantities such as gas density, temperature and velocity gradient. Radiative feedback on the ambient gas is ensured by the radiative force in the conservation of gas momentum and by the heating/cooling terms in the conservation of gas energy. In the radiative energy equation, the advection term is handled thanks to the high order approximate Riemann solvers already available in \texttt{MPI-AMRVAC} \citep{Porth14,Xia2017}. Photon-tiring is added as an explicit source term while heating and cooling are computed in an implicit way. The diffusive term is treated with the multigrid solver based on a Gauss-Seidel iterative relaxation method introduced in \cite{Teunissen2019}. This module is fully compatible with the multi-dimensional block-based adaptive mesh refinement at the basis of the domain decomposition strategy of \texttt{MPI-AMRVAC}, which enables MPI-parallelization up to an arbitrary high number of cores. It performs well on a variety of testcases. Precursors and realistic shock thickness are retrieved in 1D setups of radiatively-dominated shocks. In optically thin environments, front shocks propagate at a speed very close to the speed of light. Galilean invariance is respected and linear damping of a radiative-hydrodynamics wave quantitatively matches the predicted behavior. 

We next applied the FLD module to the launching of a radiatively-driven optically thick wind from the hydrostatic core of a Wolf-Rayet star, using a superposition of the standard OPAL opacity tables used in hydrostatics and a simple parametrisation of the significantly enhanced line-opacity expected in a supersonic outflow. In agreement with the results obtained by \cite{Poniatowski2020}, we find the OPAL opacity to be decisive in the deep and optically thick layers of the star (initiating the supersonic outflow from the so-called "iron-opacity bump" at $T \approx 2 \times 10^5$ K), while the line-opacity mechanism takes over in the outer wind, preventing the flow from falling back by bringing the outflow above the local escape speed. 

We note, however, that it is far from clear that this is what would really happen in a multi-dimensional and time-variable Wolf-Rayet outflow; indeed, in order to make the purely 1D stellar outflow escape we had to make an \textit{ad-hoc} assumption that the line-force in the outer wind is enhanced above the value expected for comparable O-stars in this region (by lowering the so-called CAK-$\alpha$ parameter, see above and also discussion in \citealt{Poniatowski2020}).
In a follow-up paper we will extend this 1D Wolf-Rayet model to 2D and 3D, in order to investigate the properties of the significant wind structure formation and time variability that presumably will occur if we instead assume more realistic conditions, and thus also allow for gas that starts to fall back upon the stellar core. The FLD code developed here is ideally suited for this project, as it is fast enough for such a multi-dimensional application while simultaneously accounting for the potential feedback from the structures on the radiative fluxes and forces. 


To this end, the FLD module has also further been tested with an an-isotropic diffusion coefficient, which might be of importance when treating line-of-sight line-opacities in a multi-D medium \citep[e.g.,][]{Kee2016}. In this formalism, the diffusion constant becomes a diagonal tensor, with the diagonal elements representing the diffusion constant in the direction of each grid line. For simplicity, however, we did not include this aspect explicitly in the paper; the reader and user can readily transform the corresponding notation in section 3.
More generally, the extension of \texttt{MPI-AMRVAC} toward general radiation(-magneto)-hydrodynamics provides us with a powerful tool suitable for a range of astrophysical applications. The FLD method is a first important step for this, and the Wolf-Rayet outflows discussed above represent a research application that can be directly considered. Another target application for FLD regards "photon-tired" very optically thick eruptive outflows from massive stars in their luminous blue variable phase (Owocki et al. 2019). Moreover, for the radiation dominated envelopes of massive stars in general, stellar models often find that the radiative acceleration exceeds gravity at the so-called "iron-opacity bump" mentioned above. It is thus possible that this, quite generally, might trigger turbulence in massive-star envelopes and atmospheres, which again might be characterised by co-existing regions of upflows and downflows (see \citealt{Jiang2015}, for some promising first simulation results). In turn, this might then provide a natural explanation for, e.g., the very broad photospheric absorption lines typically observed for O-stars, which strongly suggests the presence of supersonic velocities already in the photosphere \citep{SimonDiaz2017}.

In this respect, we plan to couple \vac to the 3D radiative transfer line-formation code by \citet{Hennicker2020}, in order to compute post-processed synthetic spectra directly from our dynamical simulations. In addition, this short-characteristics code will provide the base for another key component of our planned future work, namely an extension of the FLD method presented here toward full radiative transfer within \vac, where the Eddington tensor can be computed from the actual RTE instead of an analytic closure relation. Here an important aspect regards careful evaluation of the analytic closure relation applied in the FLD method for various regimes, as well as critical testing (and extension) of the simple line-opacity formalism for supersonic flows described in the previous section.  

\begin{acknowledgements} 
NM and JS acknowledge support by the Belgian Research Foundation Flanders (FWO) Odysseus program under grant number G0H9218N. LP and JS acknowledge support from the KU Leuven C1 grant MAESTRO C16/17/007. IEM has received funding from the Research Foundation Flanders (FWO), from the European Union's Horizon 2020 research and innovation program under the Marie Sk\l odowska-Curie grant agreement No 665501 and from the European Research Council (ERC) under the European Union’s Horizon 2020 research and innovation programme (grant agreement No 863412). JT was supported by postdoctoral fellowship 12Q6117N from Research Foundation -- Flanders (FWO). RK received funding from the European Research Council (ERC) under the European Union’s Horizon 2020 research and innovation programme (grant agreement No. 833251 PROMINENT ERC-ADG 2018).
\end{acknowledgements}

\bibliographystyle{aa}
\bibliography{main.bib}

\section*{Appendix A: Pseudo-planar correction}\label{App: pseudoplanar}
Since the multigrid method implemented for the FLD module is not capable of solving the Helmholtz equation on spherical meshes, the full system of PDEs \eqref{eq:hd_rho}, \eqref{eq:hd_mom}, \eqref{eq:hd_e} and \eqref{Eq:0th_cmf} is solved on a Cartesian grid. This means that for spherical problems, the advection terms have to be modified for spherical fluxes. In the type of simulation presented in section \ref{sec:StellarAppl}, the calculations are therefore done on a hybrid Cartesian/spherical pseudo-planar grid as presented by \citet{Sundqvist2018}. This will allow for 1D, 2D and even 3D simulations of radially extended systems on a Cartesian grid. In a 1D setting, the pseudo planar geometry is equivalent to the $r$-direction of a spherical geometry. In 2D or 3D, the $x$-direction of a pseudo planar geometry plays the role of the $r$-direction of a spherical geometry. If we neglect curvature effects i.e. if the lateral extension of the slab is small compared to its extension along $x$, then the fluxes along the lateral $y$ and $z$ directions do not require any correction.
To illustrate the method, we consider scalar conservation equations such as for \eqref{eq:hd_rho}, \eqref{eq:hd_e} and \eqref{Eq:0th_cmf}. The divergence of the vector $\vec{f}_u$ in the conservation equation for the conserved quantity $u$ contains a term with partial derivatives in the $x$-direction in Cartesian coordinates, that we note $\nabla_x \cdot (\vec{f})$, and in the $r$-direction in spherical coordinates, $\nabla_r \cdot (\vec{f})$, respectively given by:
\begin{align}
\nabla_x \cdot \vec{f}_u &= \partial_x f_{u,x}\\
\nabla_r \cdot \vec{f}_u &= \frac{1}{r^2} \partial_r (r^2 f_{u,r}) = \partial_r f_{u,r} + \frac{2f_{u,r}}{r}. 
\end{align}
We can thus assume $x\sim r$ provided we account for a geometric source term $S^g_u=-2f_{u,r}/r$. The conservation equation on a spherical grid can now be re-written as the conservation equation on a Cartesian grid plus this geometric source term:
\begin{equation}
    \partial_t u + \nabla_{x} \cdot \vec{f}_u = S^g_u.
\end{equation}
So, for density, gas energy and radiation energy:
\begin{align}
    S^g_\rho &= -\frac{2 \rho v_x}{x} \\
    S^g_e &=  -\frac{2 (e + p) v_x}{x} \\
    S^g_E &= -\frac{2 (E v_x + F_x)}{x}.
\end{align}
For the evolution equation of a vector-like conserved variable, such as momentum, the pseudo-planar correction is different as we work with the divergence of a tensor instead of a vector. In the pseudo planar approach, $\theta\sim\pi/2$ since we work near the equatorial plane of the spherical coordinate system. The coordinates $y$ and $z$ are locally equivalent to $r\theta$ and $r\phi$. The correction term can then be calculated for each of the spatial components of the momentum equation. Due to how the divergence of a tensor is defined, the correction term is different for lateral components:
\begin{equation}
   S^g_{\rho v_y} = -\frac{3 (\rho v_x v_y)}{x} ;\; S^g_{\rho v_z} = -\frac{3 (\rho v_x v_z)}{x},
\end{equation}
as compared to a radial component:
\begin{equation}
    S^g_{\rho v_x} = -\frac{2 \rho v_x^2}{x} + \frac{2 \rho v_y^2}{x}  + \frac{2 \rho v_z^2}{x}.
\end{equation}
Finally, the radiation work term in equation \eqref{Eq:0th_cmf} also features a divergence operator, so this one too needs to be corrected. This additional geometric correction source term in the radiation energy equation is:

\begin{equation}
    S^{g,\text{rad. work}}_E = \frac{2 v_x \vec{P}_{xx}}{x}.
\end{equation}






\end{document}